\documentclass{article}
\usepackage{bm}
\usepackage{amsmath,amsfonts}

\newcommand{\eq}[1]{(\ref{#1})}

\newcommand{\matnorm}[1]{{\left\vert\kern-0.25ex\left\vert\kern-0.25ex\left\vert #1 \right\vert\kern-0.25ex\right\vert\kern-0.25ex\right\vert}}
\newcommand{\eucmatnorm}[1]{{\left\vert\kern-0.25ex\left\vert\kern-0.25ex\left\vert #1 \right\vert\kern-0.25ex\right\vert\kern-0.25ex\right\vert}_{2}}

\newcommand{\RdimVec}[1]{\mathbb{R}^{#1}}

\newcommand{\Rdim}[2]{\mathbb{R}^{#1 \times #2}}

\newcommand{\innerprod}[2]{\langle #1, #2 \rangle}

\newcommand{\seq}[2]{\{#1_{k}\}_{k \in #2}}

\newcommand{\ltwo}{\ell^{2}(\mathbb{N})}
\newcommand{\lltwo}[1]{\ell^{2}(#1)}
\def\H{\mathcal{H}}
\def\N{\mathbb{N}}
\def\a{\alpha}
\def\Id{\mathcal{I}}
\def\K{\mathcal{K}}

\def\L{\mathcal{L}}
\def\Lc{\mathcal{L}^{*}}
\def\Proj{\mathcal{P}}
\def\ProjM{\mathbf{P}}
\def\PProj{\widetilde{\mathcal{P}}}
\def\PProjM{\widetilde{\mathbf{P}}}
\def\NN{\mathcal{N}}

\def\Mop{\mathcal{M}}

\def\mop{m}

\def\Mpop{\mathcal{\widetilde{M}}}
\def\Mpmat{\mathbf{\widetilde{M}}}
\def\mpop{\widetilde{m}}

\def\Tps{^{\intercal}}

\usepackage{spconf}
\usepackage{amsmath,amssymb}
\usepackage{graphicx}
\usepackage{psfrag}
\usepackage{auto-pst-pdf}
\usepackage{cases}
\usepackage{soulutf8}
\usepackage{color}
\usepackage{caption}
\usepackage{subcaption}
\graphicspath{{figures/}}

\title{THE PSEUDO PROJECTION OPERATOR: APPLICATIONS OF DEEP LEARNING TO PROJECTION BASED FILTERING IN NON-TRIVIAL FREQUENCY REGIMES}
\name{Matthew L. Weiss$\star$
    ~~~Nathan C. Frey$\star$
	~~~Siddharth Samsi$\star$
	~~~Randy C. Paffenroth$\ddagger$
    ~~~Vijay Gadepally$\star\dagger$
\thanks{Research was sponsored by the United States Air Force Research Laboratory and the United States Air Force Artificial Intelligence Accelerator and was accomplished under Cooperative Agreement Number FA8750-19-2-1000. The views and conclusions contained in this document are those of the authors and should not be interpreted as representing the official policies, either expressed or implied, of the United States Air Force or the U.S. Government. The U.S. Government is authorized to reproduce and distribute reprints for Government purposes notwithstanding any copyright notation herein.}
}
\address{
$\star$ MIT Lincoln Laboratory, Lincoln Laboratory Supercomputing Center\\
$\dagger$ MIT Connection Science\\
$\ddagger$ Worcester Polytechnic Institute, Department of Mathematical Sciences
}

\begin{document}
\ninept
\maketitle

\begin{abstract}
Traditional frequency based projection filters, or projection operators (PO), separate signal and noise through a series of transformations which remove frequencies where noise is present.  However, this technique relies on a priori knowledge of what frequencies contain signal and noise and that these frequencies do not overlap, which is difficult to achieve in practice.  To address these issues, we introduce a PO-neural network hybrid model, the Pseudo Projection Operator (PPO), which leverages a neural network to perform frequency selection.  We compare the filtering capabilities of a PPO, PO, and denoising autoencoder (DAE) on the University of Rochester Multi-Modal Music Performance Dataset  with a variety of added noise types.  In the majority of experiments, the PPO outperforms both the PO and DAE.  Based upon these results, we suggest future application of the PPO to filtering problems in the physical and biological sciences.
\end{abstract}

\begin{keywords}
\textbf{Deep Learning, Hilbert Space, Filtering, Fourier Transform, Signal Processing}
\end{keywords}

\section{Introduction} \label{sec:introduction}

In this paper, we focus on frequency based projection filters, or projection operators (PO) \cite{helmberg2008introduction,kennedy2013hilbert,vetterli2014foundations}, in the context where the frequencies corresponding to noise are unknown and/or there is non-trivial overlap between the signal and noise frequencies.  Frequency based filters are contrasted with statistically based filters, such as the Kolmogorov-Wiener filter and family of Kalman Filters \cite{anderson2012optimal,catlin2012estimation,ruymgaart2013mathematics,chang2016applied}, which assume some statistical relation between the signal and noise.  

Projection filters are based upon the idea of mapping from an abstract Hilbert Space $\H$ (e.g. the time domain) to the Hilbert Space $\lltwo{\K}$ (e.g. the frequency domain), where $\K$ is a finite or countably infinite set, via an analysis operator.  In $\lltwo{\K}$, the subspaces, or frequencies, where noise is present are weighted by 0 and the signal subspaces are weighted by 1.  The subspaces weighted by 1 are then combined via a synthesis operator, while those weighted by zero are ignored.  This results in a filtered version of the original signal.  These types of filters include the well known family of Fourier Transform based filters \cite{vetterli2014foundations}.  

One common property among projection based filters is that the selection of frequencies is rule based and, to function well, some \textit{a priori} knowledge of where signal and noise live in $\lltwo{\K}$ is required.  Furthermore, the signal and noise frequencies must have little, if any, overlap, which is difficult to achieve in complex noise environments.  To address this difficulty, we present a novel filtering algorithm, the Pseudo Projection Operator (PPO), which leverages the theoretical foundation of traditional filtering techniques in Hilbert Spaces and neural networks to perform frequency selection.  
However, in the case of the PPO, the task of frequency selection is no longer simply selecting which frequencies to keep and which to discard, but reformulated as one of weighting frequencies, where the weighting is determined by a neural network.

Prior work combining neural networks and Fourier Transforms include the composition of Fourier Transforms in hidden layers of a neural network \cite{convolutional2008cetotti}, using neural networks to learn operator solutions to partial differential equations \cite{li2020fourier-short}, substituting a neural network in place of the Fourier Transform for autonomous driving applications \cite{lopez2021spiking}, and the replacement of a self-attention layer in a Transformer with a Fourier Transform \cite{lee2021fnet}.  Furthermore, there exists work merging neural networks and statistically based filtering techniques \cite{weiss2019deep,weiss2019autoencoder,vedula2020maneuvering}.  While all the above works leverage both neural networks and some form of filtering, to our knowledge, \textit{neural networks have not been merged with traditional projection based filtering techniques to specifically address the problem of filtering in non-trivial signal and noise frequency regimes.}

\section{Theory} \label{sec:theory}
%
%
As a note to the reader, herein Hilbert Space refers to separable Hilbert Spaces and projection operator refers to orthogonal projection operators.  The basis of signal processing in Hilbert Spaces is the fact that all general Hilbert Spaces $\H$, are isomorphic to the Hilbert Space $\ltwo$, where $\N$ is the natural numbers \cite{helmberg2008introduction}.  In this context, filtering in Hilbert Spaces can be understood as selecting a subset of the sequence $\alpha = \seq{\a}{\N} \in \ltwo$.  

The isomorphism mentioned above is facilitated by two linear operators, $\Lc$ and $\L$, which are referred to as the analysis and synthesis operators respectively \cite{vetterli2014foundations}.  More formally,
\begin{align}
  \Lc : \H \mapsto \ltwo \label{eq:ana_op} \\
  \L : \ltwo \mapsto \H \label{eq:syn_op}
\end{align}
If the basis associated with \eq{eq:ana_op} and \eq{eq:syn_op} is a complete orthonormal basis $\seq{\varphi}{\N}$, the analysis operator acting on $f \in \H$ results in $\a$ and its corresponding coefficients
\begin{align}
  \a &= \Lc f \label{eq:ana_1} \\
  \a_{k} &= \innerprod{f}{\varphi_{k}} \label{eq:ana_2}
\end{align}
\noindent where $\innerprod{\cdot}{\cdot}$ is the inner product \cite{helmberg2008introduction}.  Similarly, the synthesis operator acting on $\a$ is given by
\begin{align}
  f &= \L \a \label{eq:syn_1} \\
  &= \sum_{k \in \N} \innerprod{f}{\varphi_{k}} \varphi_{k} \label{eq:syn_2}
\end{align}
\noindent Substituting \eq{eq:ana_1} into \eq{eq:syn_1} we see the composition of \eq{eq:ana_op} and \eq{eq:syn_op} forms the identity operator
\begin{equation}
  \Id = \L \Lc \label{eq:iden_op}
\end{equation}

In the context of filtering, we want something other than the identity operator in order to remove unwanted noise.  This is accomplished by introducing an operator $\Mop : \ltwo \mapsto \ltwo$ and composing it with \eq{eq:ana_op} and \eq{eq:syn_op} which results in a PO
\begin{equation}
  \Proj = \L \Mop \Lc \label{eq:proj_op} \\
\end{equation}
\noindent where $\Mop$ weights each coefficient in $\seq{\alpha}{\N}$ by \textit{either 0 or 1}.  For a signal $f \in \H$
\begin{equation}
  \Proj f =  \sum_{k \in \N} \mop_{k} \innerprod{f}{\varphi_{k}} \varphi_{k} \label{eq:proj_op_sum}
\end{equation}
\noindent with $\mop_{k} \in \{0,1\}$.  

However, we can see from \eq{eq:proj_op_sum} that $\Proj$ either removes or keeps coefficients as it weights each by either 0 or 1.  Since the coefficients \eq{eq:ana_2} weight each frequency $k$, in order for efficient filtering to occur with a PO the frequencies corresponding to noise must be known or well estimated and signal and noise need to correspond to distinct coefficients.  Here we are interested in the more general case where these criteria are not met.  To this end we introduce a new operator $\Mpop : \ltwo \mapsto \ltwo$, analogous to $\Mop$.  However, there is an important difference.  Applying $\Mpop$ to $\alpha$ weights each coefficient \textit{between 0 and 1}.  This is the essence of the PPO.  \textit{If a coefficient corresponds to only noise it can be weighted by 0, if it corresponds only to signal it can be weighted by 1, and if signal and noise are mixed that coefficient can be weighted between 0 and 1}.

The PPO and its action on $f \in \H$ are
\begin{align}
  \PProj &= \L \Mpop \Lc \label{eq:pproj_op} \\
  \PProj f &=  \sum_{k \in \N} \mpop_{k} \innerprod{f}{\varphi_{k}} \varphi_{k} \label{eq:pproj_op_sum}
\end{align}
\noindent where $\mpop_{k} \in [0,1]$.  For an operator on $\H$ to be a projection operator it must be both idempotent and self-adjoint \cite{helmberg2008introduction}.  While \eq{eq:pproj_op} is self-adjoint, it is not necessarily idempotent.  This is the reasoning behind the term ``pseudo''.  

%
%
In practice the operators introduced above are implemented as matrices.  In the case of the analysis and synthesis operators, their implementations are orthogonal matrices constructed from families of complete orthogonal functions.  Some of the more common of these families are the Legendre Polynomials, Laguerre Polynomials, Hermite Polynomials, and trigonometric polynomials \cite{lebedev1972special,powell1981approximation}.  Our implementation uses trigonometric polynomials.  

Given $x \in \RdimVec{M}$ and an index set $I_{N} = \{ 0,1,\dots,N \}$, for some integer $N$, the trigonometric polynomials which form the basis $\{ \varphi_{2k}, \varphi_{2k+1} \}_{k \in I_{N}}$ are defined as
\begin{align}
\varphi_{2k} &= \cos(kx) / \sqrt{\pi} \label{eq:trig_poly_cos} \\
\varphi_{2k+1} &= \sin(kx) / \sqrt{\pi} \label{eq:trig_poly_sin} 
\end{align}
\noindent where each basis function is in $\RdimVec{M}$ and $\sqrt{\pi}$ ensures the basis functions of orthonormal.  Stacking \eq{eq:trig_poly_cos} and \eq{eq:trig_poly_sin} as the columns of a matrix $\Phi \in \Rdim{M}{2N+2}$, $\Phi$ then correspond to the synthesis operator in \eq{eq:syn_1} and its transpose, $\Phi^{\Tps}$, corresponds to the analysis operator in \eq{eq:ana_1}.  The matrix equivalents of \eq{eq:proj_op} and \eq{eq:pproj_op} are diagonal matrices $\ProjM$ with diagonal elements in $\{0,1\}$ and $\PProjM$ with diagonal elements in $[0,1]$ respectively.  The corresponding summations in both \eq{eq:proj_op_sum} and \eq{eq:pproj_op_sum} are over $k \in \{0,1,\dots,2N+1\}$.

\section{Algorithm} \label{sec:algorithm}
\begin{figure}[t]
  \centering 
  \psfrag{1}[][l][1]{$f$}
  \psfrag{2}[][l][1]{$\Phi^{\Tps}$}
  \psfrag{3}[][l][1]{$\NN$}
  \psfrag{4}[][l][1]{$\Mpmat \Phi^{\Tps} f$}
  \psfrag{5}[][l][1]{$\Phi$}
  \psfrag{6}[][l][1]{$\PProjM f$}
  \psfrag{a}[][l][1]{$(a)$}
  \psfrag{b}[][l][1]{$(b)$}
  \psfrag{c}[][l][1]{$(c)$}
  \psfrag{d}[][l][1]{$(d)$}
  \includegraphics[scale=0.35]{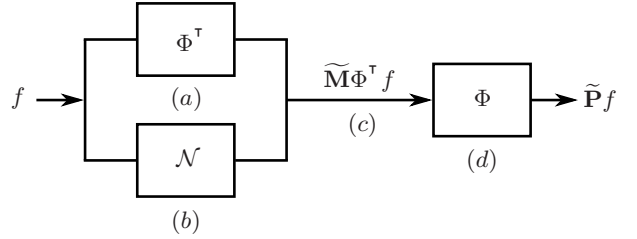} 
  \caption{The input to the PPO algorithm is a signal $f$, which is passed to two functions.  The first, represented by (a), corresponds to the analysis operator.  It is important to note that $\Phi$, and hence $\Phi^{\Tps}$, is computed offline and is not learned as part of the algorithm.  (b) represents the feed forward neural network which learns the matrix $\Mpmat$, where the activation function in the final layer of $\NN$ is a sigmoid.  Next, $\Mpmat$ acts on the output of (a) resulting in the weighted coefficients.  These are then synthesized via $\Phi$.  The final output is $\PProjM f$.}
  \label{fig:nn-diagram}
\end{figure}
As mentioned above, the goal of the PPO is to find an optimal weighting of coefficients.  To this end, we train a feed forward neural network $\NN$ to learn this weighting.  More precisely, we leverage $\NN$ to learn a diagonal matrix $\Mpmat$, corresponding to $\Mpop$ in \eq{eq:pproj_op}, where each element along the diagonal is in $[0,1]$.  To achieve this, the activation function in the last layer of $\NN$ is a sigmoid function.  The entire process of the PPO is depicted in Fig. \ref{fig:nn-diagram}.

\section{Experiments} \label{sec:experiments}

We test the PPO against a PO and a denoising autoencoder (DAE) \cite{goodfellow2016deep}. Experiments were conducted using the MIT SuperCloud~\cite{reuther2018interactive-short}.   Recall that the motivation for developing the PPO is to transfer the selection of coefficients in $\lltwo{\N}$ from a rule-based approach to one where a neural network weights the coefficients.  This is the reasoning for comparing the PPO to a PO.  Both the PPO and PO can also be thought of as learning the representation of an input $f \in \H$ in $\lltwo{\N}$.  From this perspective, the analysis and synthesis operators act as an encoder and decoder respectively.  Thus, the comparison with a DAE is included as well.  

\subsection{URMP Dataset} \label{subsec:experiments:urmp-dataset}

Experiments were conducted using the University of Rochester Multi-Modal Music Performance (URMP) Dataset \cite{li2019creating}.  This dataset consists of multi-instrument recordings of musical pieces where each instrument is recorded on an individual track.  Our process for breaking up a single recording into sub-sections is described in the context of a specific musical piece, ``The Art of the Fuguge'' (piece number 32 in the dataset), focusing on a single instrument.  This piece consists of two violins, a single viola, and single cello and runs for 2:54.  Using the cello track as our example, this track is broken into one second intervals consisting of 48,000 samples each.  This is then down sampled by selecting every $10^{th}$ sample.  The final form of this track has dimensions $\Rdim{174}{4800}$, 174 seconds and 4800 samples.  This process is applied to all individual instrument tracks, for all 44 musical pieces in the dataset, resulting in a final dataset size of $\Rdim{16,478}{4800}$.  Note, for each musical piece there is a track which mixes all individual instrument tracks which we do not include in the dataset.

The data preprocessing pipeline was as follows.  First, an 80/20 train/evaluation split was performed using Scikit Learn's ShuffleSplit class.  This is followed by scaling the dataset using either the MinMaxScaler or StandardScaler classes from Scikit Learn.  For the MinMaxScaler the feature\_range parameter was (-1,1) and for the StandardScaler the default parameters were used \cite{scikit-learn-short}.  The last step consisted of adding noise to the dataset (if applicable) and is described below.

To facilitate a variety of experimental environments we generate three types of datasets: clean, shuffle, and outliers.  Apart from the clean data, all datasets involve passing a noised version of the original data as input to a model and then optimizing (in the case of the PPO and DAE) on the clean, or denoised, data.  For all models, evaluation was performed by passing a noised input and comparing the model's output to the true, denoised data, where the evaluation metric in both training and evaluation was MSE.

The clean dataset consists of the original URMP recordings, without any added noise, discussed above.  The shuffle dataset is generated by adding a shuffled version of the clean dataset to itself.  This was included to create a situation where signal and noise frequencies overlapped.  Since many of the frequencies present in the musical pieces share frequencies, there were very likely cases where this overlap occurred.  The outliers dataset is generated by randomly scaling each of the 4800 samples in each of the 16,478 trials of the clean dataset by a real number uniformly sampled from $[1,2)$, and was included to ensure the models were trained and evaluated on noise which was a function of, or correlated with, the signal.

Here it is important to point out that, for the PPO and DAE, the shuffle and outliers datasets were resampled each epoch during training.  Thus, during training the same noised data was, effectively, never seen more than once.  In the case of the noised and denoised data used for model evaluation, these datasets were fixed, never seen during training, and written to disk to ensure all models were evaluated on the same dataset.

\subsection{Models} \label{subsec:experiments:models}

Both the PPO and DAE had three variants corresponding to the number of mappings.  For the PPO the model notation is PPOx, where x corresponds to the number of mappings in $\NN$.  The DAE followed similar notation where x represents the number of mappings in both the encoder and decoder portions.  Given the input dimensions of 4800, each PPO model mapped from $\RdimVec{4800} \mapsto \RdimVec{2048}$.  Here 2048 is the number of coefficients in $\lltwo{\N}$ (1024 coefficients for both $\sin$ and $\cos$) resulting in $N=1023$ from Section \ref{sec:theory}.  PPO3 had the mappings $4800 \mapsto 3200 \mapsto 2432 \mapsto 2048$, PPO2 dropped the 3200 dimension, and PPO1 mapped directly from 4800 to 2048.  The three DAE models, DAE1, DAE2, and DAE3, followed a similar pattern with the caveat the decoder reversed the mapping as well.  The output layer of all DAE models was an affine transformation to ensure negative values were possible.  Lastly, the PO models selected which coefficients to keep based upon a threshold.  For example, the notation PO 0.5 indicates only coefficients with an absolute value greater than to 0.5 were kept.  PO without a trailing number indicates an all pass PO, i.e. no coefficients were discarded.  In the case of the PPO and DAE, the largest model size was three as in both cases models with more than three layers did not show any improvement.  Similarly, PO models with a cutoff above 0.5 showed no improved performance.

\subsection{Results} \label{subsec:experiments:results}
\begin{figure*}[t]
    \centering
    \begin{subfigure}[t]{\textwidth}
        \centering 
        \includegraphics[scale=0.42]{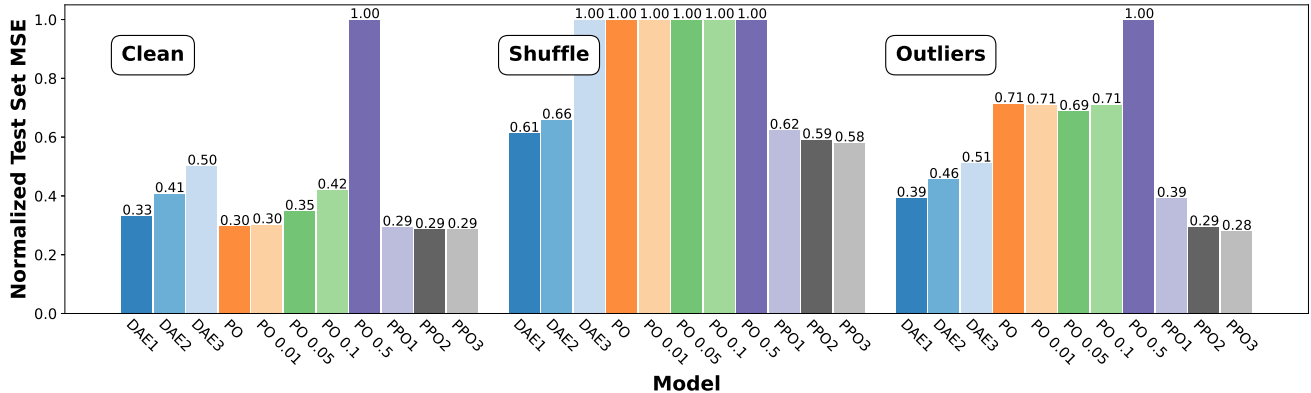} 
        \caption{Normalized evaluation set MSE for all models on all datasets with standardization scaling.}
        \label{fig:std_results}
        \end{subfigure}
        \\
        \begin{subfigure}[t]{\textwidth}
            \centering    
            \includegraphics[scale=0.42]{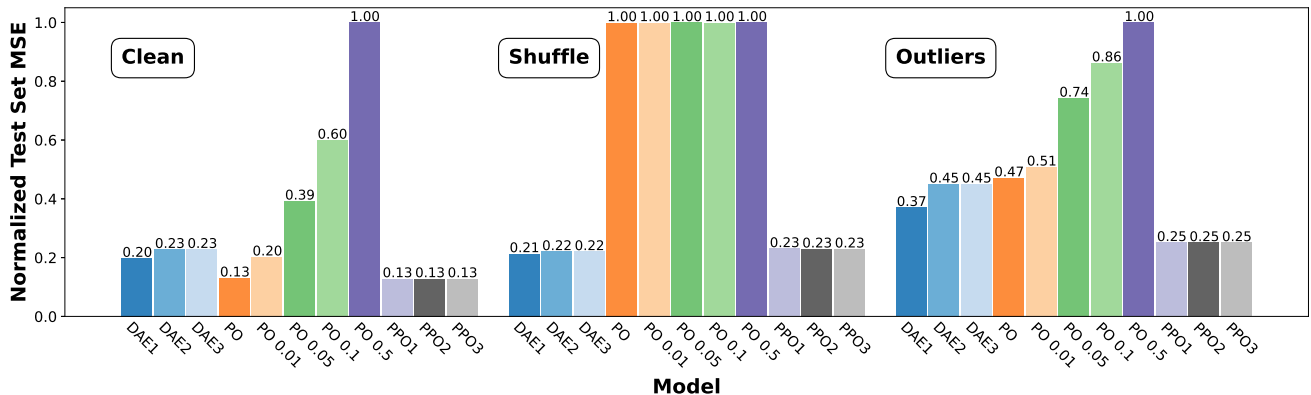} 
            \caption{Normalized evaluation set MSE for models on all datasets with min-max scaling.}
            \label{fig:minmax_results}
    \end{subfigure}
    \caption{Normalized evaluation set MSE for standardization scaling (Fig. \ref{fig:std_results}) and min-max scaling (Fig. \ref{fig:minmax_results}).  The number following DAE indicates the number of mappings in both the encoder and decoder while the number following PPO (our proposed algorithm) indicates the number of mappings in the neural network $\NN$.  The number following PO indicates all coefficients with absolute value less than that number were removed, with the caveat that PO without a trailing number indicates no coefficients were removed, i.e. an all pass filter.}
    \label{fig:results}
\end{figure*}

All PPO and DAE models were trained for 2000 epochs, with ReLU as the hidden layer activation function, L2 regularization, a batch size of 256, and over a variety of learning rates.  To prevent data snooping, for each PPO and DAE model, the optimal learning rate was selected based upon the validation set MSE and not the evaluation set MSE, where the validation set was generated using the validation\_split parameter in the tf.keras.Model.fit method \cite{tensorflow2015-whitepaper-short}.  All results in Figs. \ref{fig:std_results} and \ref{fig:minmax_results} are normalized to $[0,1]$ by dividing each dataset-scaling group by the largest value in that group. 

For the standardized clean and shuffle datasets in Fig. \ref{fig:std_results} the best performing PPO and DAE differed by only a few percentage points.  However, in the case of the standardized outliers dataset, we see the best PPO model outperformed the best DAE model by over 10\%.  In the case of min-max scaling in Fig. \ref{fig:minmax_results}, the best performing PPO outperforms the best DAE by 7\% on the clean dataset and both the PPO and DAE show insignificant differences on the shuffle dataset.  In the case of the outliers dataset, the PPO outperforms the DAE by over 10\%.  The above PPO results on the clean dataset are significant as they indicate the neural network portion of the PPO `got out of the way' when there was no added noise and the PPO is able to do as well or better than a standard autoencoder \cite{goodfellow2016deep}.  We note the DAE results in Fig. \ref{fig:results} indicate that the DAE did no better, or got worse, as the number of layers increased.  We suspect this is the result of overfitting, which may be particular to the current dataset and/or noise types.  Based on these results, we draw the following conclusions:
\begin{itemize}
\item The PPO's neural network portion is able to preserve the entire signal as well as an all pass PO when given noiseless data.
\item In the presence of random outliers, generated by scaling the dataset itself, the PPO model is optimal.
\item In the case of adding noise by shuffling the dataset the PPO and DAE models perform similarly.
\item Rule based PO filters do not do well in regimes where the added noise is random as above.
\item These results indicate a potential application of the PPO to filtering problems with partially known noise signatures which are correlated with the underlying signal as discussed below.
\end{itemize}
The problem of de-noising time series data is ubiquitous across the sciences and we envision that PPO may be applied to a variety of scientific signal processing problems. In materials science, time series data is generated during monitoring of crystal growth. The signal corresponds to the desired product, while noise is introduced by the presence of defects in the material, contaminants, and competing phases. The complexity of the growth process and the high-dimensional space of noise sources makes it difficult to establish general, rule-based filters for processing data collected during crystal growth
\cite{lee2000using-short}.  Similarly, time series data of chemical concentrations during reactions are used to determine reaction mechanisms and improve yields \cite{samoilov2001deduction}. In bioinformatics, time series are applied for metabolomics to identify the small molecules present in biological systems \cite{xia2011metatt}. These are only a few representative examples where human-expert systems and heuristic rules are used to analyze time series data to study physical and living systems, and may be augmented by the PPO.

\section{Summary and Future Work} \label{sec:summary}

In this paper we demonstrated the Pseudo Projection Operator performs as expected in the presence of denoised data, while outperforming traditional rule based POs and DAEs when the added noise was a function of the noiseless signal.  In the case where the added noise was a shuffled version of the noiseless data, the PPO and DAE performed similarly with both outperforming the PO.  Based upon this we conclude the PPO family of models would be best suited for problems where filtering is necessary in the presence of random outliers which have some correlation with the noiseless signal.  More generally, these results indicate, in the majority of cases, the PPO outperforms a DAE and PO when the noise frequencies present are unknown and/or overlap with signal frequencies.

We plan to build upon the results herein by performing time complexity and energy consumption comparisons between the PPO and DAE models, determining whether the PPO is correctly scaling coefficients in experiments where the frequencies present in the clean data are known, and in applications to filtering problems in the physical and biological sciences discussed in Section \ref{subsec:experiments:results}.

\section{Acknowledgements}
The authors acknowledge the MIT SuperCloud team: William Arcand, David Bestor, William Bergeron, Chansup Byun, Matthew Hubbell, Michael Houle, Mike Jones, Jeremy Kepner, Anna Klein, Peter Michaleas, Lauren Milechin, Julie Mullen, Andrew Prout, Albert Reuther, Antonio Rosa, and Charles Yee. The authors also wish to acknowledge the following individuals for their contributions and support: Bob Bond, Allan Vanterpool, Tucker Hamilton, Jeff Gottschalk, Tim Kraska, Mike Kanaan, Charles Leiserson, Dave Martinez, John Radovan, Steve Rejto, Daniela Rus, Marc Zissman. 

\bibliographystyle{IEEEbib}
\bibliography{main}

\begin{thebibliography}{10}

\bibitem{helmberg2008introduction}
G.~Helmberg,
\newblock {\em Introduction to Spectral Theory in Hilbert Space},
\newblock Dover Books on Mathematics. Dover Publications, 2008.

\bibitem{kennedy2013hilbert}
R.A. Kennedy and P.~Sadeghi,
\newblock {\em Hilbert Space Methods in Signal Processing},
\newblock Hilbert Space Methods in Signal Processing. Cambridge University
  Press, 2013.

\bibitem{vetterli2014foundations}
M.~Vetterli, J.~Kova{\v{c}}evi{\'c}, and V.K. Goyal,
\newblock {\em Foundations of Signal Processing},
\newblock Cambridge University Press, 2014.

\bibitem{anderson2012optimal}
B.D.O. Anderson and J.B. Moore,
\newblock {\em Optimal Filtering},
\newblock Dover Books on Electrical Engineering. Dover Publications, 2012.

\bibitem{catlin2012estimation}
D.E. Catlin,
\newblock {\em Estimation, Control, and the Discrete Kalman Filter},
\newblock Applied Mathematical Sciences. Springer New York, 2012.

\bibitem{ruymgaart2013mathematics}
P.A. Ruymgaart and T.T. Soong,
\newblock {\em Mathematics of Kalman-Bucy Filtering},
\newblock Springer Series in Information Sciences. Springer Berlin Heidelberg,
  2013.

\bibitem{chang2016applied}
C.B. Chang and K.P. Dunn,
\newblock {\em Applied State Estimation and Association},
\newblock MIT Lincoln Laboratory Series. MIT Press, 2016.

\bibitem{convolutional2008cetotti}
Hubert Cecotti and Axel Graeser,
\newblock ``Convolutional neural network with embedded fourier transform for
  eeg classification,''
\newblock in {\em 2008 19th International Conference on Pattern Recognition},
  2008, pp. 1--4.

\bibitem{li2020fourier-short}
Zongyi Li~\textit{et al}.,
\newblock ``Fourier neural operator for parametric partial differential
  equations,''
\newblock {\em arXiv preprint arXiv:2010.08895}, 2020.

\bibitem{lopez2021spiking}
Javier L{\'o}pez-Randulfe, Tobias Duswald, Zhenshan Bing, and Alois Knoll,
\newblock ``Spiking neural network for fourier transform and object detection
  for automotive radar,''
\newblock {\em Frontiers in Neurorobotics}, vol. 15, 2021.

\bibitem{lee2021fnet}
James Lee-Thorp, Joshua Ainslie, Ilya Eckstein, and Santiago Ontanon,
\newblock ``Fnet: Mixing tokens with fourier transforms,''
\newblock {\em arXiv preprint arXiv:2105.03824}, 2021.

\bibitem{weiss2019deep}
Matthew Weiss, Randy~C Paffenroth, Jacob~R Whitehill, and Joshua~R Uzarski,
\newblock ``Deep learning with domain randomization for optimal filtering,''
\newblock in {\em 2019 18th IEEE International Conference On Machine Learning
  And Applications (ICMLA)}. IEEE, 2019, pp. 1779--1786.

\bibitem{weiss2019autoencoder}
Matthew~L Weiss, Randy~C Paffenroth, and Joshua~R Uzarski,
\newblock ``The autoencoder-kalman filter: Theory and practice,''
\newblock in {\em 2019 53rd Asilomar Conference on Signals, Systems, and
  Computers}. IEEE, 2019, pp. 2176--2179.

\bibitem{vedula2020maneuvering}
Kirty Vedula, Matthew~L Weiss, Randy~C Paffenroth, Joshua~R Uzarski, and
  D~Richard Brown,
\newblock ``Maneuvering target tracking using the autoencoder-interacting
  multiple model filter,''
\newblock in {\em 2020 54th Asilomar Conference on Signals, Systems, and
  Computers}. IEEE, 2020, pp. 1512--1517.

\bibitem{lebedev1972special}
N.N. Lebedev and R.A. Silverman,
\newblock {\em Special Functions and Their Applications},
\newblock Dover Books on Mathematics. Dover Publications, 1972.

\bibitem{powell1981approximation}
M.J.D. Powell,
\newblock {\em Approximation Theory and Methods},
\newblock Cambridge University Press, 1981.

\bibitem{goodfellow2016deep}
Ian Goodfellow, Yoshua Bengio, and Aaron Courville,
\newblock {\em Deep learning},
\newblock MIT press, 2016.

\bibitem{reuther2018interactive-short}
Albert Reuther~\textit{et al}.,
\newblock ``Interactive supercomputing on 40,000 cores for machine learning and
  data analysis,''
\newblock in {\em 2018 IEEE High Performance extreme Computing Conference
  (HPEC)}. IEEE, 2018, pp. 1--6.

\bibitem{li2019creating}
Bochen Li, Xinzhao Liu, Karthik Dinesh, Zhiyao Duan, and Gaurav Sharma,
\newblock ``Creating a multitrack classical music performance dataset for
  multimodal music analysis: Challenges, insights, and applications,''
\newblock {\em IEEE Transactions on Multimedia}, vol. 21, no. 2, pp. 522--535,
  2019.

\bibitem{scikit-learn-short}
F.~Pedregosa~\textit{et al}.,
\newblock ``Scikit-learn: Machine learning in {P}ython,''
\newblock {\em Journal of Machine Learning Research}, vol. 12, pp. 2825--2830,
  2011.

\bibitem{tensorflow2015-whitepaper-short}
Mart\'{\i}n Abadi~\textit{et al}.,
\newblock ``{TensorFlow}: Large-scale machine learning on heterogeneous
  systems,'' 2015,
\newblock Software available from tensorflow.org.

\bibitem{lee2000using-short}
Kyeong~K Lee~\textit{et al}.,
\newblock ``Using neural networks to construct models of the molecular beam
  epitaxy process,''
\newblock {\em IEEE Transactions on Semiconductor Manufacturing}, vol. 13, no.
  1, pp. 34--45, 2000.

\bibitem{samoilov2001deduction}
Michael Samoilov, Adam Arkin, and John Ross,
\newblock ``On the deduction of chemical reaction pathways from measurements of
  time series of concentrations,''
\newblock {\em Chaos: An Interdisciplinary Journal of Nonlinear Science}, vol.
  11, no. 1, pp. 108--114, 2001.

\bibitem{xia2011metatt}
Jianguo Xia, Igor~V Sinelnikov, and David~S Wishart,
\newblock ``Metatt: a web-based metabolomics tool for analyzing time-series and
  two-factor datasets,''
\newblock {\em Bioinformatics}, vol. 27, no. 17, pp. 2455--2456, 2011.

\end{thebibliography}

\end{document}